# Towards High-Value Datasets determination for data-driven development: a systematic literature review

Anastasija Nikiforova[1], Nina Rizun[2], Magdalena Ciesielska[2], Charalampos Alexopoulos[3], Andrea Miletič[4]

[1] University of Tartu,

[2] Gdańsk University of Technology,

[3] University of the Aegean,

[4] University of Zagreb

Corresponding author: Anastasija Nikiforova, Anastasija.Nikiforova@ut.ee

**Abstract.** The OGD is seen as a political and socio-economic phenomenon that promises to promote civic engagement and stimulate public sector innovations in various areas of public life. To bring the expected benefits, data must be reused and transformed into value-added products or services. This, in turn, sets another precondition for data that are expected to not only be available and comply with open data principles, but also be of value, i.e., of interest for reuse by the end-user. This refers to the notion of "high-value dataset" (HVD), recognized by the European Data Portal as a key trend in the OGD area in 2022. While there is a progress in this direction, e.g., the Open Data Directive, incl. identifying 6 key categories, a list of HVDs and arrangements for their publication and re-use, they can be seen as "core" / "base" datasets aimed at increasing interoperability of public sector data with a high priority, contributing to the development of a more mature OGD initiative. Depending on the specifics of a region and country – geographical location, social, environmental, economic issues, cultural characteristics, (under)developed sectors and market specificities, more datasets can be recognized as of high value for a particular country. However, there is no standardized approach to assist chief data officers in this. In this paper, we present a systematic review of existing literature on the HVD determination, which is expected to form an initial knowledge base for this process, incl. used approaches and indicators to determine them, data, stakeholders.

## 1 Introduction

Open Government Data (OGD) is seen as an emerging political and socio-economic phenomenon that promises to benefit the economy, improve the transparency, efficiency, and quality of public services, including the transformation of government data-driven actions, stimulate public sector innovations in various areas of public life and promote civic engagement [1,2]. OGD is considered to have a positive impact on the lives of individuals / citizens, and society, and contributes to tackling environmental problems, contributing to efficient, data-driven, sustainability-oriented development. In other words, OGD is considered to have economic, social, and environmental value. OGD is also one of the pillars of Open Government, which in turn is seen to be supported not only by OGD availability, but by the availability of high-value OGD.

However, the value of data, and open government data in particular, is a complex topic. In other words, although "value" itself is seen as a multifaceted concept, if the user of the data is clearly known, it can be determined from that user's viewpoint/ perspective, i.e., needs, expectations and understanding of "value", i.e., what this concept means for this particular user. However, OGD, by definition, does not have a predetermined user – OGD are freely available, accessible and provided without restrictions for their further reuse, which means that they can be used by everyone, regardless of age, gender, education, specialization, and for the purpose they will find necessary for themselves, which, as practice shows, in many cases differs significantly from the original purpose of collecting and using data internally by data producers / owners. The latter, in turn, is seen as a key determinant of the success of OGD as a movement, philosophy and policy, where data are provided for their further re-use and transformation into value (incl. services, products, new business models) when various forms of intelligence, incl. data intelligence, artificial intelligence, embodied intelligence, and collaborative intelligence [3], are used as a method for their transformation.

This makes the determination of OGD value a difficult task, and while several attempts have been made to date, there is considerable room for research and improvement for both academia and public administration. This is all the more so with the reference to the determining "high-value datasets" (HVD), particularly being interested in facilitating data provision, i.e., identifying potentially valuable datasets that are not yet available / open, i.e., when *ex-post* assessment of how valuable, interesting and / or useful was the dataset in question is not possible.

In terms of current progress in this area, this refers to a list of initiatives and studies carried out by several organizations and communities, where at the European level, probably most notable progress has been made by the European Commission in the Open Data Directive (originally Public Sector Information Directive (PSI Directive) in relation to the notion of HVD, referring to **datasets whose re-use is expected to create the most value for society, the economy, and the environment, contributing to the creation of "*value-added services, applications and new, high-quality and decent jobs, and of the number of potential beneficiaries of the value-added services and applications based on those datasets*"** [4]. To date, an agreement on six HVD thematic categories was reached as part of *Directive (EU) 2019/1024 of the European Parliament and of the Council of 20 June 2019 on open data and the re-use of public sector information*, according to which there are six thematic data categories of HVD – (1) geospatial, (2) earth observation and environment, (3) meteorological, (4) statistics, (5) companies and company ownership, (6) mobility data are considered as of high value [4].

Further, a list of specific HVDs and the arrangements for their publication was developed and made available as "*Commission Implementing Regulation (EU) 2023/138 of 21 December 2022 laying down a list of specific high-value datasets and the arrangements for their publication and re-use*" [5] that can be seen as seeking for greater harmonization and interoperability of public sector data and data sharing across EU countries with reference to specific datasets, their granularity, key attributes, geographic coverage, requirements for their re-use, including licence (Creative Commons BY 4.0, any equivalent, or less restrictive open licence), specific format where appropriate, frequency of updates and timeliness, availability in machine-readable format, accessibility via API and bulk download, supported with metadata describing the data within the scope of the INSPIRE data themes that shall contain specific minimum set of the required metadata elements, description of the data structure and semantics, the use of controlled vocabularies and taxonomies (if relevant) etc. In addition, the Semantic Interoperability Community (SEMIC) is constantly hosting webinars on DCAT-AP (Data Catalogue Vocabulary Application Profile) for HVD to discuss with OGD portal owners, OGD publishers and enthusiasts the best approaches to use DCAT-AP to describe HVD and ensure their further findability, accessibility, and reusability.

As a follow-up to this initiative, the Open Data Maturity Report 2022 (ODMR2022) has updated their methodology with a focus on countries' readiness for HVD and the upcoming European Commission Implementing Act, adding relevant indicators to all four dimensions it covers (policy, impact, portal, and quality), which are measured using a questionnaire answered by a representative of the open data initiative in question, asking about the preparatory measures that countries might have initiated before the Implementing Act and the actions taken to date. Thus, our study can be seen as a response to this call seeking to help countries take these steps, especially given the very fragmented understanding of the topic in the scientific literature and lack of connection between public administration and academia in this question.

The recent ODMR2022 also reported that most countries are taking at least some steps to identify HVD and take related activities, and some of them perform well, but HVD list identification is based on categories prioritised at EU level. In other words while it can be seen that progress has been made in this area, an examination of the above documents reveals that these datasets rather form a list of "mandatory" or "open by default" datasets, sometimes also referred to as "base" or "core" datasets, aiming at open data interoperability with a high level of priority and a relatively equal level of value for most countries, which contributed to the development and promotion of a more mature open data ecosystem (ODE) and OGD initiative.

However, the value of data is known to depend on a perspective such as the user's point of view, where [6], for instance, highlights the need to consider both the perspective of the data publisher and the perspective of data re-user when evaluating the value of data and defining HVD. In addition, the value of data depends on country-specific aspects, such as geographic location and its specificities, current environment, social, economic issues, culture, ethnicity, likelihood of crises and / or catastrophes, (under)developed industries/ sectors and market specificities, and development trajectories, i.e., priorities. Depending on the above, more datasets can be recognized as having high value within a particular country or region [7-9]. For example, meteorological data describing sea level rise can be of great value in the Netherlands as it has a strong impact on citizens and businesses as more than 1/3 of the country is below sea level, however, the same data will be less valuable for less affected to countries, such as Italy and France [7]. We believe that additional factors such as ongoing smart cities initiatives, as well as the Sustainable Development Goals, the current state of countries and cities in relation to their implementation and established priorities affect this list as well.

Therefore, it is important to support the identification of country specific HVD that will increase user interest by transforming data into innovative solution and services. Although this fact is recognized by countries and some local and regional efforts, mostly undertaken by governments with little support from the scientific and academia community, they are mainly faced with problems in the form of delays in their development or complete failure, or ending up with some set of HVD, but little information about how this was actually done. These ad-hoc attempts remain closed and not reusable, which is contrary to both the general OGD philosophy and the HVD-centric philosophy that is expected to be standardized. Most of them are *ex-post* or a combination of the *ex-ante* and *ex-post*, making the process of identifying them more resource-intensive, with an effect only visible after potentially valuable datasets have been discovered, published, and kept maintained, with the need for further evaluation of their impact, which is a resource-consuming task. It is also in line with [10], according to which there is no standardized approach to assisting chief data officers in identifying HVDs, resulting in a failure in consistent identification and maintenance of HVDs.

Considering the importance of this topic, it makes sense to refer to the literature and summarize what has been conducted so far in this regard. Thus, the objective is to examine how HVD determination has been reflected in the literature over the years and what has been found by these studies to date, incl. the indicators used in them, involved stakeholders, data-related aspects, and frameworks. This is done by conducting a Systematic Literature Review (SLR). To achieve the research objective, the following research questions (RQ) were established:

- *(RQ1) how is the value of the open government data perceived / defined? In which contexts has the topic of HVD been investigated by previous research (e.g., research disciplines, countries)? Are local efforts being made at the country levels to identify the datasets that provide the most value to stakeholders of the local open data ecosystem?*
- *(RQ1.1) how the high-value data are defined, if this definition differs from the definition introduced in the PSI /OD Directive, and (RQ1.2) what datasets are considered to be of higher value in terms of data nature, data type, data format, data dynamism?*
- *(RQ2) What indicators are used to determine high-value datasets? How can these indicators be classified? Can they be measured? And whether this can be done (semi-)automatically?*

- *(RQ3) Whether there is a framework for determining country specific HVD? In other words, is it possible to determine what datasets are of particular value and interest for their further reuse and value creation, taking into account the specificities of the country under consideration, e.g., culture, geography, ethnicity, likelihood of crises and/or catastrophes.*

While the results are expected to be of greater interest to public administration and public agencies to understand which datasets are most in demand with their subsequent opening (or academia to set the research agenda), the current state of affairs and current trends in a broader open data ecosystem perspective suggests that these findings can be of interest to a wider audience, with respect to B2G, C2G and other data governance models. The results are also expected to form the knowledge base for the framework for determining HVD, while the validation of identified indicators (as part of this study and derived from government reports) takes place during the workshops with open (government) data and / or e-government experts (2 editions have already taken place).

The paper is structured as follows: Section 2 provides an overview of the topic, emphasizing the importance of HVD, Section 3 presents the methodology of this study, Section 4 provides the results, Section 5 establishes a discussion around the findings, and provides key conclusions and directions for further research.

## 2  Background

The practice shows that there is often a mismatch between users' needs and the opportunities offered by available datasets [11], where OGD are not relevant to the problems that users want to solve [12]. In other words, open data is often found to be either superficial or irrelevant to potential users [2]. However, while the use and value of OGD are often discussed, encouraging data agencies to open their data, governments to support the opening and maintenance of data, citizens, businesses and other data user groups to use these data, there are relatively few studies that analyse the actual use and value of data (and specific dataset), particularly without limiting the scope for a subset of data. This is due to the complexity of the topic and the underlying general paradigm of the OGD – freely accessible and freely reusable, i.e., the license allows to reuse data without declaring it in the resulting product or service does, there are a limited number of indicators and approaches to measure the use and the value of data, especially the value of an individual dataset, as well as the success of data opening policies.

Those studies that make such an attempt, typically address the use and value of government data from one of two perspectives – (1) qualitative, (2) quantitative [2]. For the qualitative perspective, motivations for using data, practices and experiences of users to reuse data and (co-)create value are typically the subject of research [2,11,13-14]. For quantitative approach, the reference to quantitative parameters such as the of views and downloads provided by the OGD portals is related to the use of datasets, although it is clear that these parameters may indicate some interest in the dataset and can be used rather as assumptions to draw onto usage trends [2,15-16], i.e., the fact that a dataset has been viewed does not guarantee that it will be actually used, where even the fact of downloading it is not a guarantee, where its actual reuse transforming it into the value is the expected end-result. For the latter, in turn, the "value" into which a dataset has been transformed, however, it is also unclear whether this transformation will have either social, economic, or environmental value itself. This makes it very challenging to estimate the value of existing datasets, not to say about those not yet published.

In 2020, the European Commission, however, attempted to address this issue, by publishing an "*Impact Assessment study on the list of High Value Datasets to be made available by the Member States under the Open Data Directive*" [17]. Based on a literature review of six thematic categories of HVD, six macro characteristics of potential value derived from open data were found, which include economic benefits, environmental benefits, generation of innovative services and innovation (innovation and artificial intelligence), reuse, improving, strengthening, and supporting public authorities in carrying out their mission. Multiple categories of value were found for each macro characteristic, resulting in a total of 32 categories of value and 126 possible indicators (both quantitative and qualitative) to measure this value. However, although it seems to be the longest list of indicators today, two workshops conducted by the authors with e-government and OGD experts found that only a few indicators are sufficiently clear, reasonable, and feasible (i.e., do not require the collection of supporting data, whose amount and complexity is higher than of original dataset).

## 3 Methodology

To understand how HVD determination has been reflected in the literature over the years and what has been found by these studies to date, we studied all relevant literature covering this topic. In order to identify relevant literature, the SLR was carried out to form the knowledge base. This was done by searching digital libraries covered by Scopus and Web of Science (WoS). Given the specificity of the topic, we covered Digital Government Research library (DGRL) that covers studies related to domains of digital government, digital governance, and digital democracy.

These databases were queried for keywords *("open data" OR "open government data") AND ("high-value data\*" OR "high value data\*")*, which were applied to the article title, keywords, and abstract to limit the number of papers to those, where these objects were primary research objects rather than mentioned in the body, e.g., as a future work. Only articles in English were considered, while in terms of scope, both journal articles, conference papers, and chapters were studied. For the period covered by these searches, we do not set nor the start date, not the end date. The query resulted in 11 articles in Scopus and 5 in WoS (Fig. 1). After deduplication, 11 articles were found unique and were further checked for relevance, all of which were found to be relevant after the first round of evaluation, and 1 article was excluded from further analysis as non-eligible. One more study was excluded because we were not able to access the full text (neither from Digital Libraries, nor from journal or conference proceedings, nor from ResearchGate). As a result, a total of 9 articles were further examined. Each study was independently examined by at least two authors.

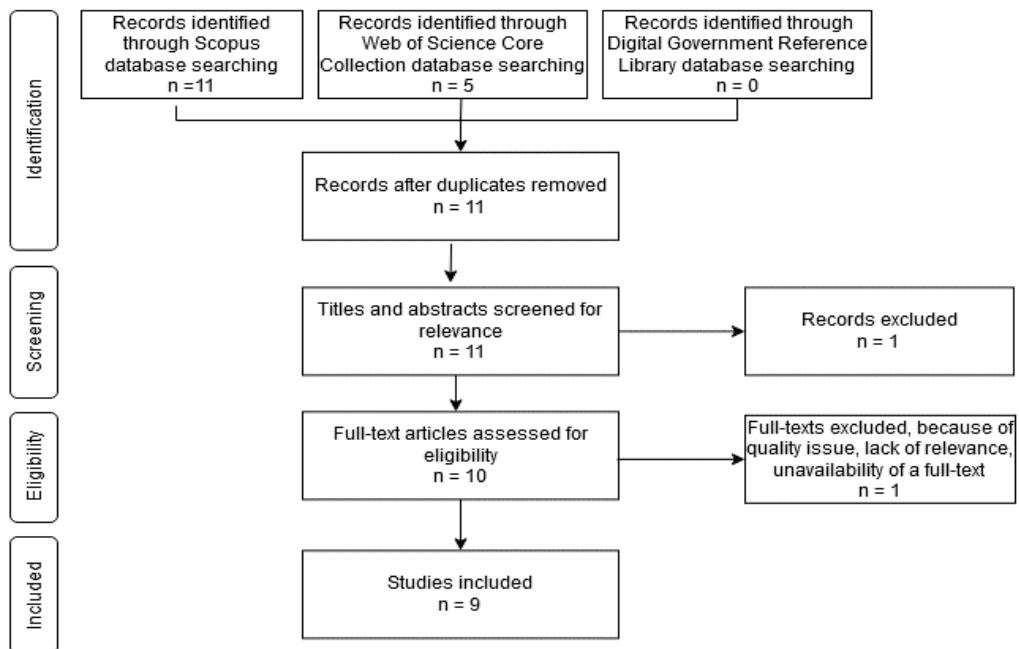

**Fig. 1.** Study selection, assessment, and inclusion (presented using the PRISMA flow diagram).

To attain the objective of our study, we developed the protocol based on [18], where the information on each selected study was collected in four categories: (1) descriptive information, (2) approach- and research design-related information, (3) quality-related information, (4) HVD determination-related information (see Table 1). The data underlying the study are publicly available on Zenodo - https://doi.org/10.5281/zenodo.7944424[1].

Table 1. The structure of the protocol.

| Category | Metadata | Description |
| --- | --- | --- |
| Descriptive information | Article number | A study number, corresponding to the study number assigned in an Excel worksheet |
| | Complete reference | The complete source information to refer to the study |
| | Year of publication | The year in which the study was published |
| | Journal article / conference paper / book chapter | The type of the paper, i.e., journal article, conference paper, or book chapter |
| | DOI / Website | A link to the website where the study can be found |
| | Number of citations | The number of citations of the article in Google Scholar, Scopus, Web of Science |
| | Availability in OA | Availability of an article in the Open Access |
| | Keywords | Keywords of the paper as indicated by the authors |
| | Relevance for this study | What is the relevance level of the article for this study? (high / medium / low) |
| Approach- and research design-related information | Objective / RQ | The research objective / aim, established research questions |
| | Research method (including unit of analysis) | The methods used to collect data, including the unit of analysis (country, organisation, specific unit that has been analysed, e.g., the number of use-cases, scope of the SLR etc.) |
| | Contributions | The contributions of the study |
| | Method | Whether the study uses a qualitative, quantitative, or mixed methods approach? |
| | Availability of the underlying research data | Whether there is a reference to the publicly available underlying research data e.g., transcriptions of interviews, collected data, or explanation why these data are not shared? |
| | Period under investigation | Period (or moment) in which the study was conducted |
| | Use of theory / theoretical concepts / approaches | Does the study mention any theory / theoretical concepts / approaches? If any theory is mentioned, how is theory used in the study? |
| Quality- and relevance- related information | Quality concerns | Whether there are any quality concerns (e.g., limited information about the research methods used)? |
| | Primary research object | Is the HVD a primary research object in the study? (primary - the paper is focused around the HVD determination, secondary - mentioned but not studied (e.g., as part of discussion, future work etc.)) |

---

[1] Anastasija Nikiforova, Nina Rizun, Magdalena Ciesielska, Charalampos Alexopoulos, & Andrea Miletič. (2023). Dataset: A Systematic Literature Review on the topic of High-value datasets [Data set]. Zenodo. https://doi.org/10.5281/zenodo.7944425

| | HVD definition and type of value | How is the HVD defined in the article and / or any other equivalent term? |
|---|---|---|
| HVD determination-related information | HVD indicators | What are the indicators to identify HVD? How were they identified? (components & relationships, "input -> output") |
| | A framework for HVD determination | Is there a framework presented for HVD identification? What components does it consist of and what are the relationships between these components? (detailed description) |
| | Stakeholders and their roles | What stakeholders or actors does HVD determination involve? What are their roles? |
| | Data | What data do HVD cover? |
| | Level (if relevant) | What is the level of the HVD determination covered in the article? (e.g., city, regional, national, international) |

## 4 Results from the Systematic Literature Review

### 4.1 Descriptive and Quality analysis

As part of the descriptive analysis, we studied the selected studies' objectives (see Table 2), the journals and conferences where these studies were published, the years of publication, the databases through which we found them, how well they were cited in these digital libraries. Most studies are exploratory and were published between 2012 and 2023. Although neither OGD, nor the importance of the value of data are new topics, scholarly publications dedicated to the topic of HVD are still limited, in contrast to OGD research in general, particularly considering that we did not apply strict exclusion criteria and did not limit the period under investigation – only two studies covered the topic in question in the last three years – [19] and [9]. The low number of relevant studies points out the limited body of knowledge on this topic, thereby making this study unique and constituting a call for action. Papers mostly come from Europe with contributions from Belgium, Bulgaria, Germany, Greece, Latvia, the Netherlands, Sweden, Switzerland, although there are several contributions coming from the United Kingdom, the United States of America, Thailand, India, and China.

**Table 2.** Overview of studies included in our systematic literature review.

| Reference | Study Objective | Primary research? |
|---|---|---|
| [20] | To develop an EnAKTing-based integrated account of how to bring OGD into the linked-data Web (LDW) | secondary |
| [21] | To develop an understanding of the various kinds of technical, administrative & regulatory challenges faced by government agencies in India while trying to release their datasets in open domain | secondary |
| [8] | To comprehensively analyse the currently available HVDs (in Thailand) in various aspects, including their domain coverage, data attributes and categorization, and intrinsic data qualities | **primary** |
| [22] | To develop a framework to assess the impact of releasing open data by applying the Social Return on Investment (SROI) approach | secondary |
| [23] | To summarize the data value evaluation methods of scientific and technological information in the open-source environment. | secondary |
| [24] | To build a bridge between linked data research and new potential adopters of the technology by providing a stepwise introduction on how to move from basic | secondary |

| | business data described in various formats housed in government registries towards linked OGD | |
|---|---|---|
| [25] | To develop a comprehensive understanding of the challenges faced by government agencies with specific focus on India while trying to release their datasets in open domain | Secondary |
| [19] | To prioritize research questions and identify community needs for data and computational simulation capabilities to foster the development of robust tools to simulate the impact of natural hazards on structures, lifelines, and communities | **? (domain-specific HVD)** |
| [9] | To identify the current value of the OGD and their compliance with the term of HVD in users' view, identifying the most valuable areas and datasets for Latvian citizens and businesses (SME) | **primary** |

For most studies HVD is a secondary research object (7 out of 9 studies), where their role differs either from an application domain, i.e., a list of already determined HVD categories and the analysis of the correspondence of datasets provided by the national OGD portal to this list [24, 8], or assessing the impact of available datasets focusing on a subset of data represented by datasets belonging to HVD [22]. In some studies, the concept of HVD is referred to more as a buzzword [20-21, 25], but it is not given enough attention, only emphasizing their importance, including their key role for the sustainability of OGD initiative.

Only two studies openly shared the underlying research data – [9,19], despite not only a growing trend towards open sharing of research data as a good open science practice, but also compliance with the general OGD philosophy and the postulated importance of their re-usability and re-use with another study that provided the reference to the data but whose maintenance was stopped and the data are no longer available. As for those studies that provided an underlying data, it should be mentioned that respective studies were published between 2021 and 2023, which can be argued as related to the growing popularity of open science and OGD as more than data, i.e., rather a philosophy and the mindset that needs to be changed as part of this movement.

For the quality analysis of the articles covered, the research design was mostly appropriate, while several studies were lacking information on approach.

### 4.2 Content analysis

#### 4.2.1. The definition of HVD

4 of 9 articles provide a clear definition of what is meant by HVD in each study, and two more studies where the definition is indirect, but at least a rough definition of such can be extracted. Nikiforova [9] uses the definition proposed in the Open Data Directive (EU) 2019/1024. Utamachant & Anutariya [8] uses the "local" definition of HVD given by the Electronic Government Agency of Thailand (EGA). Wang et al. [23] suggests a high-level definition, according to which HVD are data that meet the actual needs. This is somewhat similar to two other studies that limit the definition of HVD to the domain they belong to, i.e., [24] limits HVD to business data only, while Zsarnóczay et al. [19] define them as *critical datasets for their work,* especially those with greater granularity and spatial extent, with the potential for re-use limited to disaster recovery planning. Stuermer & Dapp [22], while not defining HVD, later in the study refer to specific data categories as defined in the G8 Open Data Charter. Alternatively, Shadbolt et al. [20] does not provide a definition of HVD, but instead postulates that understanding of HVD will be missing until existing data is available in a linked data web (LDW), which will help understand the demand side and collect relevant feedback for local, regional, and national. Since some studies instead of defining HVD have focused more on data-related aspects, let us refer to them.]

A total of 5 studies reflected on HVD as data categories or specific datasets (Table 2). 2 of them refer to already existing classifications such as the 14 categories defined by the G8 Open Data Charter [22], the 9 HVD categories defined by the EGA [8], while [9] uses a list of 6 HVD data categories defined by the OD Directive to design a questionnaire for citizens and SMEs to identify HVD from their point of view, coming up with 9 categories that were named as HVDs by respondents in addition to the OD Directive (some of which are rather subsets of these categories). Following a similar survey and consultation approach, [19] does not use any predefined classification, identified 4 HVD categories with 12 subcategories for the area of disaster recovery planning.

An analysis of the literature as well as prior work in this area allows us to group these data-related aspects into (1) data categories, (2) specific datasets, (3) data type, (4) data dynamism. For the latter groups, they can be seen in part as prerequisites for being considered truly valuable and potentially reusable, mentioning the need for increased interoperability, including providing these data in LOGD (Linked Open Government Data) form (the importance of RDF and LOGD is emphasized in [20,24]. In other cases, it refers to certain features that today make data more prospective for transforming into value-adding products and services, with geospatial data, real-time data and sensor-generated data being predominantly mentioned. Otherwise, some studies, especially more recent ones, are increasingly mentioning datasets that are related to or share the same values as Sustainable Development Goals, the concept of Smart Cities, where both data generated as part of their operation and data that can potentially contribute to the development and maintenance of a smart city are seen as HVD. And yet another category that, though not so often mentioned now, certainly deserves attention, is the emphasis placed by several studies on data that can be seen as citizen-generated.

It should be emphasized that although most studies we have reviewed are of national level conducted in Latvia, Greece, Thailand, India, UK, USA, only two of them actually focus on country-specific HVD, namely Latvia [9] and Thailand [8], where the second rather evaluates the list of datasets denoted as HVD on the OGD portal, while [19] focuses on the determination of HVD in a specific domain. This means that determination of country specific HVDs in the scientific literature is very underrepresented.

Otherwise, the vast majority of research perceive HVD as data that will be of interest to be re-used, where the type of value created by this re-use may differ, as well as the beneficiary (business, citizen, government, or both). Thus, let us now discuss what stakeholders and actors these studies cover or consider important.

Table 2. Overview of data categories and datasets recognized as HVD

| Ref. | Data categories or datasets |
| --- | --- |
| [8] | **9 HVD data categories defined by EGA**<br>(1) Politics and government, (2) government budget and spendings, (3) economic, financial and industry, (4) public healthcare, (5) law, justice and crime, (6) social and welfare, (7) agriculture and irrigation, (8) art, culture and religion, (9) ICT and communication; 22 datasets out of more than 1000 |
| [20] | **14 HVD categories defined by the G8 Open Data Charter**<br>(1) Companies, (2) Crime and Justice, (3) earth observations, (4) education, (5) energy and environment, (6) finance and contracts, (7) geospatial, (8) global development, (9) Government Accountability and Democracy, (10) Health, (11) Science and Research, (12) Statistics, (13) Social mobility and welfare, (14) Transport and Infrastructure |
| [24] | **1 pre-defined category**<br>Business data, e.g., basic data about a company (e.g., legal name, address, representative, establishment date and company type), company identifiers and annual balance sheets. Authoritative data, LOGD |
| [19] | **4 categories (12 subcategories) defined by respondents**<br>(1) Buildings; (2) Households, Businesses, and Services; (3) Recovery; (4) Hazard |
| [9] | **6 HVD categories defined by the OD Directive**<br>(1) geospatial data, (2) earth observation and environment, (3) meteorological, (4) statistics, (5) companies and company ownership, (6) mobility; |

| | **9 categories defined by the respondents:** (1) medical and health data, (2) detailed tourism data on regions, (3) transport data, (4) data on the suitability of transport and places for people with disabilities, (5) data on streets and traffic lights, (6) data on radiation and noise levels, (7) data on physical and mental health of people, (8) data on social media (e.g., hash tags, fake news, data leaks), (9) sensor data |

### 4.2.2. Which stakeholders does HVD determination involve?

Most studies mention general ODE stakeholders, with only a few focused on those related to the definition of HVD. More precisely, Nikiforova [9] conducts a survey of Latvian citizens / society (individuals) and businesses, in particular small and medium-sized enterprises (SME) since the definition of HVD should take into account the experience of industry reusers, who are characterized by a deeper understanding of this "value", knowing what kind of data may be needed for a particular application. Zsarnóczay et al. [19], conduct the workshops with researchers, developers, and practitioners with expertise in the domain they study, i.e., earthquake, coastal, and wind hazards from engineering, planning, data sciences, and social science, some of whom are also data providers. Their rationale is similar to the above, i.e., the input should be obtained from real users, considering their needs and, as a result having a demand to be reused (corresponds to [6,26]).

Utamachant & Anutariya [8], who, however, assess the compliance of already published HVD to their perception of HVD, while mentioning that the European data portal recommends that two different points of view be considered when identifying HVD – the data provider and the data re-user, does not follow this recommendation. They argue that OGD public participation in Thailand is still at an immature stage and determining a community data demands is almost impossible, leading to the need to use an alternative approach to identify HVD by referring to the world standards, such as those defined by Government Open Data Index (GODI) and Open Data Barometer (ODB), with further examination of the list by the domain experts, who in turn evaluate the impact of these datasets and readiness for their opening. This, i.e., immaturity of community engagement, should be highlighted as a barrier not only to the HVD determination, but to the overall success of a healthy and sustainable OGD ecosystem. Similarly, Shadbolt et al. [22] argue that OGD is not a rigid government IT specification – it requires a productive dialogue between data providers, users, and developers, where a "perpetual beta" should be expected, in which best practice, technical development, innovative use of data, and citizen-centric policies come together to drive data release programs.

As regards the general ODE stakeholders these studies mention, which is not surprising considering that they are all ultimately part of the HVD publishing process, while those listed above are seen the ones who should identify them as potential reusers. [9] also pointed out that the literature suggests that determining the value of a particular dataset is a very complex and multi-perspective task, when the data provider, who tends to have data usage statistics such as views, downloads and number of showcases / use-cases / re-uses seen as one of the popular and most widely used HVD indicators [6,7,15], plays a critical role, which is also because there are different views on who should benefit from HVD and an impact they create, i.e., HVD value beneficiary.

Otherwise, two general categories of OGD stakeholders are (1) data producers, also called data publishers, data providers, data suppliers, who in the traditional OGD ecosystem are government and public agencies, and (2) data consumers, also called data users, reusers, who can be then broken down into smaller groups such as citizens, society, NGO, developers / innovators, entrepreneurs, business and SME in particular, start-ups, media, data journalists, researchers / scientists / academic community, domain experts, private and public sector etc. Government and public agencies should also be considered policy makers, and, more importantly, OGD users, while the rapid changes of the OGD ecosystem suggest that all those listed as data users may also act as data providers, thereby more frequently viewing the OGD ecosystem as a combination of various data governance models that are no longer limited to G2C or G2B, with B2G, C2G becoming increasingly popular as an integral part of the OGD ecosystem rather than an independent, or at least the need for these changes is postulated.

**4.2.3. HVD determination: indicators and a framework**

As mentioned before only 2 (3) studies address HVD as a primary topic, and even fewer studies focusing on the process of determining HVD. However, we should mention that while the above represents the so-called "ex ante" approach, some studies, on the contrary, refer to what is called "ex post", i.e., proposing methods along with indicators for measuring actual impact and value of data. And although they make little contribution to the determination of HVD, they can be seen as a valuable asset in the process of opening and maintaining HVD. In other words, the data are expected not only become publicly available in an open data format, but their actual impact and value must also be measured. These outputs, in turn, may provide some insight into the determination of the next set of HVD, based on public interest in a given category of data or, in contrast to very low resuse rates, although there is evidence that a particular topic of data is in a great demand.

In more detail, given the impossibility of reaching the wider community in Thailand, [8] takes an alternative approach that we would rather define as verification of the HVD datasets determined by their OGD portal. The study proposes an approach to identify HVDs among already published datasets. To this end, they map the datasets denoted as HVD to the Government Open Data Index (GODI) and Open Data Barometer (ODB) data categories, which are then filtered using the UK National Information Infrastructure (NII) datasets, which are then processed by domain experts. Initially, Corruption Perception Index (CPI) was also used along GODI and ODB.

Shadbolt et al. [20], however, consider that such a proposal would only be possible when LDW is ensured, which would help to identify HVD. However, there is no clear idea of how exactly this is supposed to be then done. Stuermer & Dapp [22] proposes a SROI-based (Social return on investment) framework to evaluate the impact of already published datasets. It consists of four values adapted to the given context, namely: (1) input - resources such as native data, money, people, infrastructure, equipment, (2) output - tangible deliverables / directly controllable results, i.e., setup and operation of an open data portal with metadata, updated content, open format etc, (3) outcome - all direct and indirect consequences of certain output actions of open data users, incl. hackathons, apps, new firms, data linking, research etc., and (4) impact – the outcome adjusted for the effects that would have occurred without the intervention, i.e., actually caused by releasing the data-value-creating consequences. These four values are linked to 14 HVD categories of the G8 Open Data Charter to create a matrix of open data examples, activities, and impacts in each data category. This can be seen as a possible asset in determining the potential of the HVD under consideration.

Wang et al. [23] do not provide a framework, but summarize the knowledge found in literature, according to which two methods are used to evaluate the OGD value, where one is based on a top-down macroeconomic method and the other is based on a bottom-up microeconomic method. A top-down macro-economic approach evaluates the total value of various industries that use OGD as an investment and shows the total value of OGD to the economy. The bottom-up microeconomic approach is concerned with analysing the productivity of OGD, taking into account the inputs and outputs of individuals and companies when using government data.

Varytimou et al. [24] suggests that the indicator to be used is a high potential for reuse in national and cross-border settings. However, they do not suggest how this high reuse potential can be determined. [19], however, considers that the most appropriate approach is to survey or interview actual real users and data providers, which, however, given the fact that the study is carried out in predefined settings (domain), where the stakeholders are more likely known being also less in nature, is easier than for the whole OGD, where the data users are generally difficult to identify.

Similarly, [9] also does not come up with a framework but believes that consultation with different stakeholder groups should be a mandatory component, where the perspectives of different stakeholder groups should be considered separately, since the value that the OGD brings to them may also differ, i.e., more economic, and entrepreneurial for SMEs and rather social and environmental for citizens. Also, for user groups such as citizens, the author suggests grouping them based on their level of familiarity with OGD and behaviour patterns to reduce the level of noise in the collected data, although considering those who have only used OGD a few times or even never, and those, who are experienced users. The results, however, are subject to interpretation, where it is important to understand what exactly refers to the data and their value, and what – to other ODE components. To structure the survey the author used the six key HVD thematic categories defined in the OD Directive (the list of datasets was not developed at that point), along with the data categories on the national OGD portal, which are evaluated for relevance, after which, when some knowledge base is created, respondents are invited to suggest their own categories, and, more importantly, specify the datasets they consider to be HVD, with further re-evaluation of their interest in OGD, if these datasets are made available. This was considered as the first step towards developing a framework, and the next phase of the analysis was expected to take place with the OGD data re-users, i.e., those who indicated themselves as re-users in the first phase and agreed to be contacted – no other approach is possible in Latvia, where OGD reuse is not monitored. This was expected to be then combined with other approaches found in the literature, incl. downloads statistics (if the holders of the Latvia Open Data Portal have these data since the portal user interface does not have this data), to ensure more complete analysis of possible indicators to be then.

The findings of the SLR and components associated with HVD determination are summarized in Figure 2, covering those HVD determination-related features we identified in the literature. Identified approaches and determinants in particular can also be divided into those of more qualitative nature and quantitative. In other words, some indicators / determinants can be used to quantify the value/ potential interest in the dataset in question, while some of them are very qualitative in nature (e.g., citizens awareness of an environmental issue, which, although can be somehow quantified, is still rather qualitative).

Depending on the source of these determinants, i.e., the input data that will be used to determine them, they can also be divided into internal – those that imply from the data the data publisher have, the owner of national open government data portal owner (e.g., Google Analytics, log files of the portal etc.), and external – where the involvement of external actors or stakeholders (both people and systems) becomes necessary to obtain the input that subsequently transforms into an understanding of potential HVDs or serves as part of the basis for such decision-making.

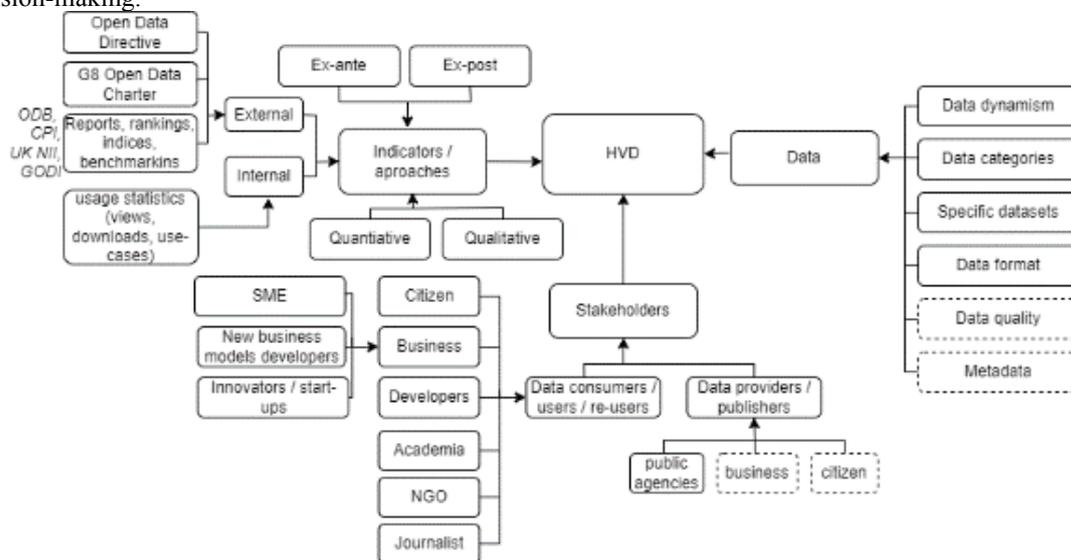

**Fig. 2.** Conceptualization of the reviewed literature around the HVD concept

We also believe that regardless of the above category, indicators can be divided into those that can be measured by quantifying this "value", and those that cannot. However, most of those indicators that have been found in the literature and Deloitte report, are neither measurable, nor SMART (specific, measurable, achievable, relevant, and time-bound), which is contrary to our expectation for the identification of such indicators. Although the objective of this study has been achieved, future work will cover examination of publicly available approaches taken by governments, extracting the indicators they use, and assessing their reliability, as well as SMARTness, which would be preferable option that will reduce the overall complexity of the HVD management (determination and maintenance) process. At the same time, it is clear that it is not always the case that non-SMART indicators are not suitable for determining HVDs. Thus, while having as a long-term objective HVD determination framework that would be based solely on SMART indicators, it is clear that they may predominate, but cannot be the only set of indicators used. What is also important to keep in mind is that the determination of HVD is not a one-time event, but a continuous process in which not only the opening / publishing and subsequent ongoing maintenance of not only the open dataset, but the entire HVD determination process as whole is expected. Thus, we can think of the HVD determination process as a lifecycle similar to the Deming, also PDCA cycle (plan-do-check-act) or define-measure-analyse-improve-control also known as phases of Lean Six Sigma, which consists of at least: identification of a list of potential HVDs, considering the current data supply and stakeholders' needs, including an analysis of the possibilities of opening HVD, opening/ publishing HVD, evaluation of an impact of a HVD in comparison with the expected, possibly considering price-to-value ratio, take a decision on the need for adjustments.

## 5 Conclusions and Future Work

The objective of this study was to examine how the HVD determination has been reflected in the literature over the years and what has been found by these studies to date, where our SLR leads us to conclude that this topic is very underrepresented in the literature. Thus, the claim that there is no standardized approach to assisting chief data officers in identifying HVDs, remain valid.

However, during this study, we have established some knowledge based on HVD determination-related aspects, including several definitions of HVD, data-related aspects, stakeholders, some indicators and approaches that can now be used as a basis for establishing a discussion of what a framework for determining HVD should look like, which, along with the input we received from a series of international workshops with open (government) data experts, covering more indicators and approaches found to be used in practice, could enrich the common understanding of the goal, thereby contributing to the next open data wave [27].